\documentclass[12pt]{article}
\usepackage{times}
\usepackage{geometry}
\usepackage[utf8]{inputenc}
\usepackage{enumitem,amssymb}
\usepackage{ragged2e}
\usepackage{setspace}
\usepackage{wrapfig}
\usepackage{graphicx}
\newlist{thematic}{itemize}{8}
\setlist[thematic]{label=$\square$}
\usepackage{pifont}
\newcommand{\cmark}{\ding{51}}%
\newcommand{\done}{\rlap{$\square$}{\raisebox{2pt}{\large\hspace{1pt}\cmark}}%
\hspace{-2.5pt}}

\newcommand\apj{The Astrophysical Journal}
\newcommand\apjl{The Astrophysical Journal, Letters}     
\newcommand\pasa{Publications of the Astronomical Society of Australia}
\newcommand\solphys{Solar Physics}
\newcommand\ssr{Space Science Reviews}

\begin{document}
\pagenumbering{gobble}
\RaggedRight
\noindent {\fontsize{16}{20} \selectfont Astro2020 Science White Paper}
\begin{center}
{\fontsize{24}{32}\selectfont Radio, Millimeter, Submillimeter Observations of the Quiet Sun}
\vspace{0.5cm}

\end{center}

\vspace{0.3cm}

\normalsize

\noindent \textbf{Thematic Areas:} \hspace*{60pt} $\square$ Planetary Systems \hspace*{10pt} $\square$ Star and Planet Formation \hspace*{20pt}\linebreak
$\square$ Formation and Evolution of Compact Objects \hspace*{31pt} $\square$ Cosmology and Fundamental Physics \linebreak
\done Stars and Stellar Evolution \hspace*{1pt} $\square$ Resolved Stellar Populations and their Environments \hspace*{40pt} \linebreak
  $\square$    Galaxy Evolution   \hspace*{45pt} $\square$             Multi-Messenger Astronomy and Astrophysics \hspace*{65pt} \linebreak

\justifying

\noindent \textbf{Principal Author:} \\
Tim Bastian, National Radio Astronomy Observatory \\
Email: tbastian@nrao.edu.edu \\
Phone: (434) 296-0348 \\

\noindent \textbf{Co-authors:} \\
Bin Chen, New Jersey Institute of Techology \\
Dale E. Gary, New Jersey Institute of Technology \\
Gregory D. Fleishman, New Jersey Institute of Technology \\
Lindsay Glesener, University of Minnesota \\
Colin Lonsdale, MIT/Haystack \\
Pascal Saint-Hilaire, University of California, Berkeley \\
Stephen M. White, Air Force Research Laboratory \\

\noindent\textbf{Executive Summary} Identification of the mechanisms responsible for heating the solar chromosphere and corona remains an outstanding problem, one of great relevance to late-type stars as well. There has been tremendous progress in the past decade, largely driven by new instruments, new observations, and sophisticated modeling efforts. Despite this progress, gaps remain. We briefly discuss the need for radio coverage of the 3D solar atmosphere and discuss the requirements. 

\pagebreak
\pagenumbering{arabic}

\section{Introduction}

An outstanding problem in solar physics and by extension, stellar physics, is how the dynamic chromosphere and corona are heated. The chromosphere and corona are only visible to the naked eye during solar eclipses, the chromosphere as brilliant, ruby-red ring of beads just above the occulted photosphere; the corona as a pearly white crown. The fundamental question is by what {\sl non-radiative} mechanism(s) the temperature of the chromosphere is heated to $>10^4$ K, and how the corona is heated to several $\times 10^6$ K. Leading theoretical ideas for how the chromosphere and corona are heated involve some form of resonant wave heating (e.g., van Ballegooijen et al. 2011) or ``nanoflares'' (Parker 1988), with many variants of these ideas under study: see, for example, recent work (Priest et al. 2018, Syntelis et al. 2019) that explores a flux cancellation nanoflare model.

This white paper summarizes progress made during the past decade, particularly new observations and progress in modeling, and points out the need for enabling the unique diagnostic potential offered by broadband radio imaging spectropolarimetry in the coming decade. 

\section{New Instrumentation, New Models}

Significant progress has been stimulated by new observations provided by a range of new observational capabilities that have come online during the past decade, with more soon to come. In addition, increasingly advanced models, motivated by new observations, have been developed with which to test our understanding. Large facilities and missions include the following:

\noindent{\bf Solar Dynamics Observatory (SDO)}: The SDO, a previous decadal priority, was launched in 2010. It carries the Helioseismic and Magnetic Imager, producing high resolution photospheric magnetograms and dopplergrams; the Extreme Ultraviolet Variability Experiment, which makes precision measurements of the UV irradiance, which heats the Earth's upper atmosphere and creates the ionosphere; and the Atmospheric Imaging Assembly, which produces high resolution images ($0.6"$/pixel)of the Sun in extreme UV wavelengths that sample temperatures from $5\times 10^4$ K to $20\times 10^6$K. 

\noindent{\bf Interface Region Imaging Spectrograph (IRIS; De Pontieu et al. 2014)}: IRIS is a small explorer that was launched in 2013. IRIS provides high resolution (0.35"), high cadence images and spectra of the solar chomosphere at UV wavelengths, sampling lines with temperatures ranging from $5000$ K to $65\times10^3$K.

\noindent{\bf Atacama Large Millimeter/Submillimeter Array (ALMA)}: Although dedicated in 2012, it was not until 2016 that ALMA became available for solar observing at mm-$\lambda$ with arcsec imaging capabilities (Shimojo et al. 2017, White et al. 2017). Emission from the quiet Sun at mm-$\lambda$ originates from the low chromosphere. The source function is Planckian and since the emission is in the Rayleigh-Jeans regime, the observed flux density is linearly proportional to the kinetic temperature of the emitting plasma. ALMA observations are therefore highly complementary to those at ultraviolet wavelengths.

\begin{figure}[htp]
\includegraphics[trim=1in 2in 1in 2in,clip,width=\textwidth] {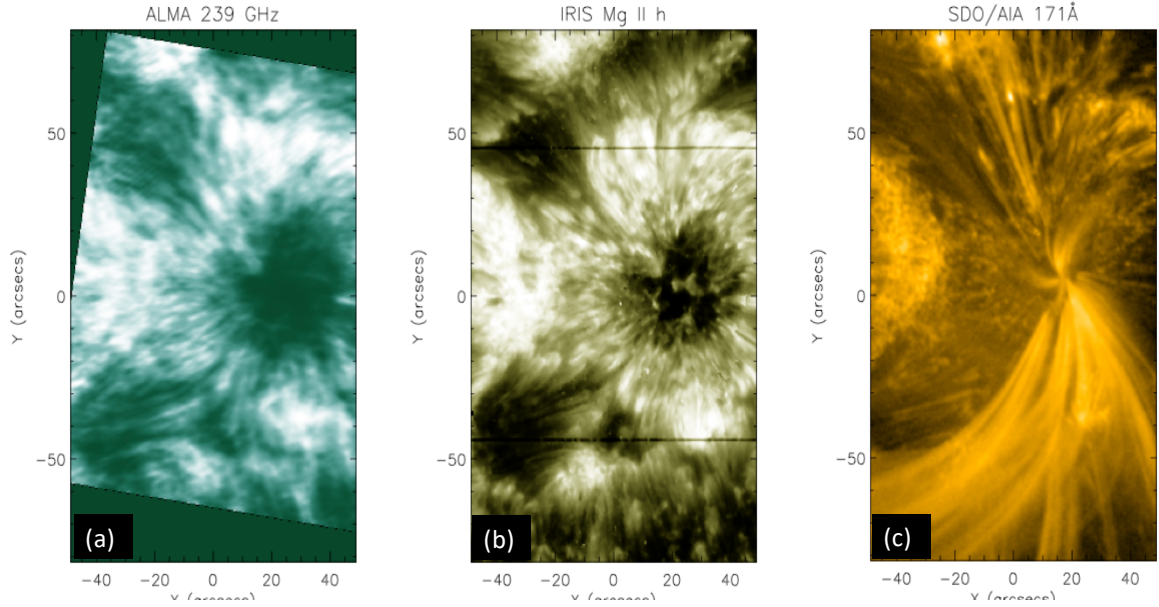}
\caption{Example of ALMA, IRIS, and SDO observations of a large sunspot. After Bastian et al. 2017.}
\label{fig:ALMA}
\end{figure}

\noindent And soon to come is:

\noindent{\bf Daniel K. Inouye Solar Telescope (DKIST)}: In 2020 DKIST will become available as the flagship ground-based solar telescope optimized for O/IR performance. DKIST will provide ultra-high resolution imaging (0.1") of the solar photosphere and chromosphere at spectral lines that span the many scale heights of the chromosphere and will resolve magnetic elements on the smallest scales. It will, moreover, advance coronal magnetography. 

These missions and facilities have been supplemented by several other ground based and sub-orbital initiatives. Examples include the Chromospheric Lyman-Alpha Spectro-Polarimeter (CLASP) program (e.g., Giono et al. 2017), the balloon-borne Sunrise experiment (Solanki et al. 2010); and the Goode Solar Telescope (Goode et al. 2010). All have opened new observing regimes that have enabled new diagnostics; and, as a result, have inevitably raised new questions. IRIS and ALMA have opened the ultraviolet (UV) and millimeter (mm) wavelength regimes for complementary high-resolution exploration of the solar chromosphere, the thin, dynamic, and complex interface between the visible photosphere and the hot solar corona. SDO has provided sustained, high-resolution and high-time-resolution records of both quiescent and active phenomena (flares, CMEs, jets, filament eruptions) in multiple passbands, providing new insights into the structure and dynamics of the thermal solar atmosphere. 

On the modeling front, significant effort has gone into advanced numerical models of the dynamic chromosphere in light of high-resolution, time-resolved observations of UV lines by IRIS, lines that form under conditions of non-LTE. These go well beyond the semi-empirical hydrostatic models that served as references for years (e.g., Veranazza, Avrett, \& Lowser 1981) to 3D radiative magnetohydrodynamic codes (e.g., the Bifrost model: Gudiksen et al. 2011; see also Carlsson et al. 2016 and references therein). For the corona and solar wind, the BATS-R-US (van der Holst 2014 and references therein) model has served as an innovation platform and reference. 

\section{A Gap in Radio Coverage}

However, despite a wealth of new observations and modeling,  fundamental questions remain: what are the dominant mechanisms that heat the solar atmosphere from chromospheric to coronal heights, what are the mechanisms of transport and loss, and how does the quiet atmosphere link to the expanding solar wind? 

\begin{wrapfigure}{r}{0.5\textwidth}
\includegraphics[trim=0in 0in 0in 0in,clip,width=3.in]{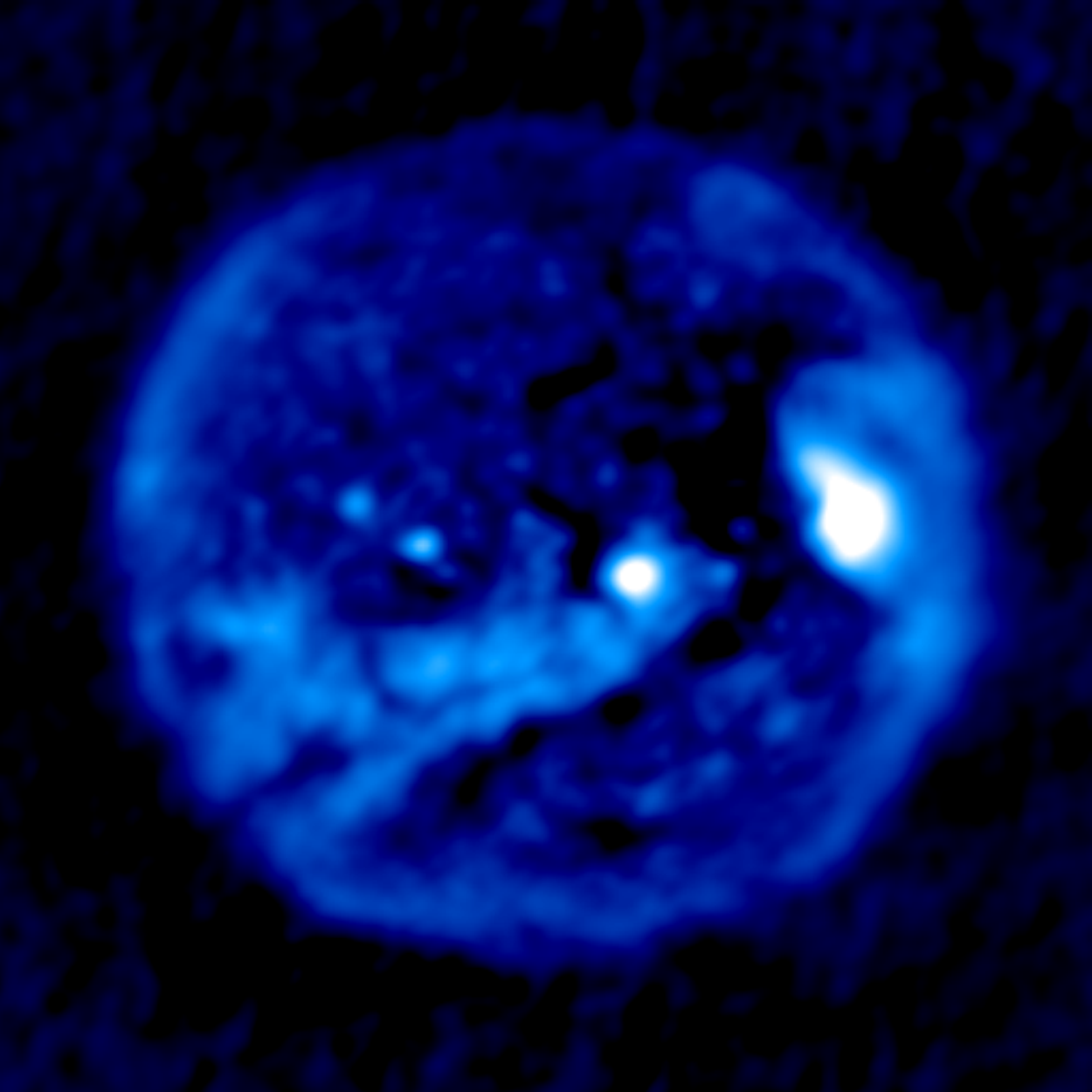}
\caption{Example of a map of the Sun at 20 cm obtained by the VLA in its D configuration, yielding an angular resolution of $\approx 40"$. Note the north polar coronal hole and its extension onto the disk.}
\end{wrapfigure}

A serious gap in observational coverage to date is the lack of sustained and comprehensive radio observations with the required frequency coverage, sensitivity, and angular resolution. Radio emission from the quiet Sun is dominated by thermal bremsstrahlung emission although nonthermal emission plays a role (microflares, jets, and other transient perturbations). In contrast to UV and EUV emission from the Sun (IRIS, SDO), the radiative transfer of radio and millimeter emission from the Sun is straightforward and serves as an important counterpoint to UV/EUV observations.\footnote{The degree of ionization can depart significantly from the expectations of statistical ionization equilibrium in the solar
chromosphere. Therefore, while the source function is in LTE, the opacity can be far from its LTE value.} The success of ALMA is notable, and it has already raised interesting challenges in this regard (e.g., Bastian et al. 2017). 

It is important to note that ALMA only probes the solar chromosphere. To realize the potential of radio diagnostics, full radio coverage from the chromosphere to the corona is needed. 
Partial coverage is provided by non-solar-dedicated instruments. The {\sl Jansky Very Large Array} provides imaging capabilities from 1-50 GHz, sampling the middle chromosphere into the low corona. Occassional mapping is performed at discrete frequencies by the VLA (e.g., Fig. 2). Work by Schonfeld et al. 2015, for example, explored the coronal sources of the 10.7 cm flux, a key proxy for EUV irradiance. But mapping of the full VLA frequency range on a regular basis is not possible. 

The Murchison Widefield Array (Tingay et al. 2013) is a low frequency imaging array operating in Australia at meter wavelengths (70-300 MHz). The Sun and heliosphere are a key component of the MWA science program. It has rejuvenated meter-wavelength observations of the Sun with new findings on radio bursts and the quiet Sun. Sharma et al. (2018) find evidence for extremely weak nonthermal emission from the quiet corona using high-dynamic-range MWA observations. Coronal holes are regions in the solar atmosphere that are magnetically opent to the solar wind. They are the source of fast solar wind streams which, in turn, can drive recurrent geomagnetic activity. The structure and connection of CH to fast streams is of key interest. Rhaman et al. (2018) have recently shown that CH emission at meter wavelengths is dark relative to background corona at a wavelength $<2$~m, yet becomes brighter at loner wavelngths. McCauley et al. (2019) show that CH emission shows fascinating polarization signatures that are not yet understood (Fig. 3). 

\begin{figure}
\begin{center}
\includegraphics[trim=2.5in 2in 2.5in 2in,clip,width=5in]{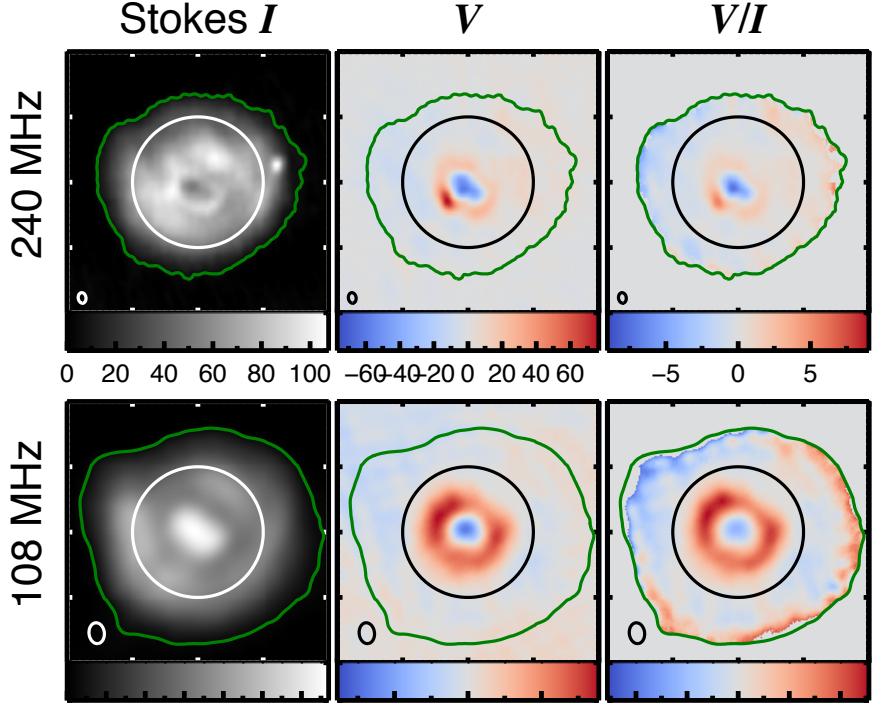}
\caption{Examples of coronal emission at meter wavelengths obtained by the MWA. The coronal near disk center is dark at 240 GHz ad bright at 108 MHz  while Stokes V is negative in the CH and positive near the boundary of the coronal hole at both frequencies. The white and black circles indicate the apparent size of the visible disk. After McCauley et al. 2019.}
\end{center}
\end{figure}

\section{Requirements for Next-Generation Radio Instrumentation}

To exploit the unique sensitivity of radio observations to both thermal and nonthermal emissions from the "quiet" solar atmosphere and to trace the solar magnetic field from chromospheric heights up into the corona, the following instrumental requirements must be fulfilled:
\begin{itemize}
\item Frequency coverage: continuous coverage from 50 MHz (6~m) to ~20 GHz (1.5 cm) to enable 3D imaging of the solar atmosphere
\item Spectral resolution: a spectral resolution of $\Delta\nu/\nu\approx 1$\% 
\item High time resolution: in order to observe weak microflares and smaller energy release events, a time resolution of 1~s is needed. 
\item High angular resolution: scattering in the Sun's corona limits the usable angular resolution to roughly $20''/\nu_9$ where $\nu_9$ s the frequency in GHz, i.e., 1'' at 20 GHz.
\item High-dynamic-range imaging in any given frequency: a dynamic range of at least $10^3:1$ is needed at each frequency. 
\item Dual-polarization performance: Measurements of the total intensity are required as are those of the Stokes V parameter, which contains quantitative information about the magnetic field. The Faraday depth of the corona is very high, washing out linearly polarized emission. 
\end{itemize}

Fulfillment of these requirements ensures that new constraints on the electron number density, temperature, and the magnetic field can be provided in 3D that are complementary to existing and planned missions and facilities. 

These requirements are met by a next-generation radioheliograph that is known as the Frequency Agile Solar Radiotelescope (FASR), a facility that has been recommended as a priority by previous decadal surveys, both the Astronomy \& Astrophysics decadals and the Solar \& Space Physics decadals. As a mid-scale-sized project, funding mechanisms have not been available to move this priority project forward. Until now. With the implementation of NSF's Mid-scale Research Infrastructure program, FASR can now be made a reality. 

\end{document}